# VehSense: Slippery Road Detection Using Smartphones

Yunfei Hou*, Abhishek Gupta†, Tong Guan†, Shaohan Hu‡, Lu Su†, and Chunming Qiao†

*School of Computer Science and Engineering, California State University, San Bernardino, U.S.
†Department of Computer Science and Engineering, University at Buffalo, U.S.
‡IBM Research, Yorktown Heights, NY, U.S.
Email: hou@csusb.edu, {agupata33, tongguan}@buffalo.edu, shaohan.hu@ibm.com, {lusu, qiao}@buffalo.edu

*Abstract*—This paper investigates a new application of vehicular sensing: detecting and reporting the slippery road conditions. We describe a system and associated algorithm to monitor vehicle skidding events using smartphones and OBD-II (On board Diagnostics) adapters. This system, which we call the VehSense, gathers data from smartphone inertial sensors and vehicle wheel speed sensors, and processes the data to monitor slippery road conditions in real-time. Specifically, two speed readings are collected: 1) ground speed, which is estimated by vehicle acceleration and rotation, and 2) wheel speed, which is retrieved from the OBD-II interface. The mismatch between these two speeds is used to infer a skidding event. Without tapping into vehicle manufactures' proprietary data (e.g., antilock braking system), VehSense is compatible with most of the passenger vehicles, and thus can be easily deployed. We evaluate our system on snow-covered roads at Buffalo, and show that it can detect vehicle skidding effectively.

## I. INTRODUCTION

Slippery road conditions during winter time, i.e., snowy, slushy, or icy pavement, are major concerns for road safety. In the U.S., winter road maintenance accounts for roughly 20% of state Department of Transportation maintenance budgets. Despite this investment, 24% of weather-related vehicle crashes occur on snowy, slushy or ice pavement, over 1,300 people are killed and more than 116,800 people are injured each year [1]. Keeping roadways in good condition is challenging because inclement weather degrades road condition just over a short period of time. Because government agencies only have limited resource, prioritizing which roads need snowplowing or deicing comes of interest. More importantly, we need to inform drivers about hazardous road conditions, especially when visibility is poor during snowfall or sleet.

To address this need, we describe the design, implementation and evaluation of VehSense, a smartphone based vehicular sensing system. Our focus in this paper is on the slippery road detection component of VehSense, we defer a discussion of our on-going project on VehSense, a comprehensive crowdsourcing platform, to future work. Particularly, VehSense uses inertial sensors (i.e., accelerometers and gyroscopes) and GPS in a smartphone, and reads from vehicle's Controller Area Network (CAN) bus (with smartphone connected with an OBD-II adapter via Bluetooth). The system relies on the mobility of vehicles to traverse roads being monitored. A smartphone based approach is a good match for road monitoring problem because: 1) it can be easily integrated with a navigation application to provide slippery road information along with other rich sensing data; 2) it is cost-effective, since sensing leverages existing sensors and communication infrastructure (although our method requires a OBD-II adapter to interface with vehicle's CAN bus, the adapter only costs about $20 dollars); and 3) it provides a reliable and systematic approach, since data are aggregated from multiple vehicles.

Only relying on the OBD-II interface to detect vehicle skidding is possible but problematic. Relevant vehicle sensor data (e.g., suspension deflection, antilock braking system and traction control) are considered proprietary by vehicle manufactures and are not available. In addition, the CAN message protocol can differ significantly by the make, model and year of vehicles, resulting in extra setting and calibration cost for different vehicles.

The intuition behind the proposed slippery road detection algorithm is simple. The smartphone fixed inside a vehicle collects 3-axis acceleration and gyroscope data at a high frequency. Meanwhile, it gathers vehicle speed data from vehicle's CAN bus via OBD-II. Each of these sensory data provides a way to get vehicle speed: 1) the acceleration data can be used to estimate the vehicle speed relative to the ground, which we refer to it as the *ground speed*; 2) the vehicle speed data reported by the wheel speed sensors (a type of tachometer from the CAN network), which we refer to it as the *wheel speed*. The key observation here is that wheel speed sensors measure the rotation speed of the wheel, however, the "actual" speed of the vehicle might be different: During normal driving conditions, there will be a good match between the ground speed and the wheel speed, but if the road is slippery, the instantaneous ground speed and wheel speed maybe different, which indicates skidding events.

This simple description hides various design challenges of the proposed VehSense for slippery road detection, we will elaborate our solutions in the following sections. The contributions of this work are:

1. To the best of our knowledge, VehSense is the first smartphone based system for detecting slippery road conditions. It utilizes low-cost, off-the-shelf OBD-II adapter. It is easy to install and does not require user calibrations. The source code of the sensing component in VehSense is available online [2]. It is capable of collecting OBD-II data at about 15 Hz (via Bluetooth with ELM 327 [3] integrated circuit).

2. We develop a skidding detection algorithm that is able to successfully detect slippery locations. Our field test at Buffalo, New York has a very low false-positive rate for skidding, flagging less than 3% of events from skidding as normal on our field test.

## II. RELATED WORK

An effective slippery road detection system is beneficial for traffic safety and road maintenance. Several vehicular sensing systems have been proposed for detecting slippery roads. These systems can be categorized into: a) image-based techniques [6]–[9], which require stereo cameras and/or infrared sensors for recording, and are computationally intensive to process, b) proprietary OBD-II based techniques [10], [11], which require proprietary data from the vehicle's antilock braking system (ABS) or ABS equivalent systems (e.g., individual wheel speed, steering angel). Vehicle manufactures generally do not allow proprietary data to be accessed by the OBD-II connection, and different manufacturers are using different CAN message sets. To the best of our knowledge, VehSense is the first system that attempts to detect slippery road condition using smartphones and is compatible with most, if not all, vehicle manufactures.

The most related project to the proposed systems is the DataProb [11] project conducted by Michigan Department of Transportation, which is based on the same hardware settings as those of VehSens. No conclusions were drawn from the DataProb project, due to limited amount of data. The DataProb project also explored the mismatch between wheel speed and ground speed as a detector for skidding, however, GPS speed reading was used as the ground speed which does not provide enough sample rate (GPS sample rate is usually up to 1 Hz) and accuracy.

## III. SYSTEM DESIGN

### A. Overview

The VehSense system consists of equipped vehicles and a central server, as illustrated in Fig. 1. Each vehicle collects the following raw information:

$$< time, location, acceleration, rotation, wheel\ speed, RPM >$$

The first four parameters come from the sensors in a smartphone: GPS provides time and location data at a rate of 1 Hz. 3-axis accelerometer and gyroscope provide acceleration and rotation data at a rate of 200 Hz. In the proposed skidding detection algorithm, acceleration and rotation data are combined to estimate the ground speed (i.e., the speed of a vehicle relative to the ground). Although GPS also provides speed information, our field test shows that GPS on the smartphone does not have the necessary precision for the slippery road detection. The last two parameters come from the OBD-II interface, wheel speed and RPM (Revolutions per Minutes, which measures the engine speed) data are collected at 10 Hz.

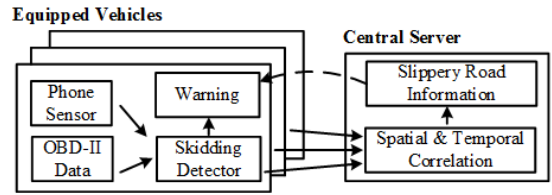

Figure 1. System overview

These sensing data streams are processed by the smartphone to produce vehicle skidding detections. The vehicles transmit collected sensor data back to the backend server through various channels, e.g., Wi-Fi or cellular network. In our preliminary work, a vehicle-to-vehicle sharing mechanism is also implemented to let two nearby vehicles share their data, thus increasing the possibility that the data reaches the central server [4]. The central server clusters skidding events reported by multiple vehicles, and applies data aggregation and fusion techniques (e.g., truth discovery), producing the output of the central server: a series of slippery road segments warnings with varying degree of possibility and severity. Finally, the slippery road warning information can be integrated into a navigation application, thus raising the awareness of potential slippery hazards and help to reduce the risk of crashing.

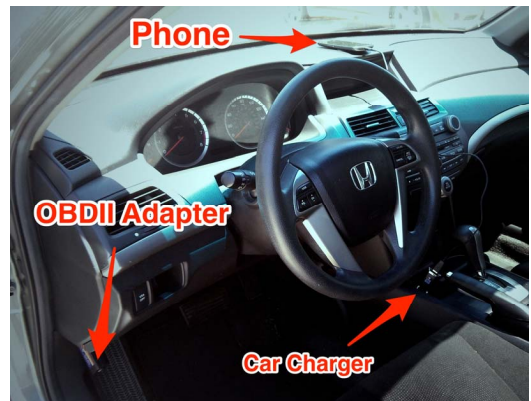

Figure 2. Equipped vehicle

## B. Vehicle Testbed

The hardware components used for equipping vehicles are: a smartphone, an OBD-II adapter and preferably, a car charger (as shown in Fig. 2). Latest version of the VehSense app runs on Android 6.0 Marshmallow and supports OBD-II adapters with ELM 327 integrated circuit (The ELM 327 integrated circuit from Elm Electronics is one of the most popular OBD-II-to-PC interpreter, and supports the standard OBD-II protocols). The smartphone can be placed at any convenient locations for the driver, as long as it is firmly attached to the vehicle.

## IV. SKIDDING DETECTION ALGORITHM

The idea behind the skidding detection algorithm is that the wheel speed (i.e., the vehicle velocity inferred by sensing the rotations of the wheel) will be significantly different from the ground speed (i.e., the actual velocity of a vehicle) caused by vehicle skidding. With VehSense, the wheel speed can be collected from OBD-II readings, at the same time, vehicle's ground speed can be inferred from smartphone sensor readings. Thus we can detect skidding conditions by analyzing the *deviation of the differences* between those two readings (i.e., a high deviation of their difference indicates slippery road surface). The major advantage of using the deviation of speed difference is that it compensates for the inherent error on calculating wheel speed and ground speed. Both of the speeds may not be the actual velocity of a vehicle: the wheel speed might suffer from errors caused by variations in tire and wheel diameter or poor calibration, while the ground speed (calculated by Equation (1)) maybe inaccurate due to accumulated error. Given this observation, we exam the variation of their difference rather than their absolute difference within a sliding time window. More specifically, the proposed algorithm works as follows:

Step 1) Preprocessing and Calibration of the Acceleration Vector

In general, the smartphone maybe in an arbitrary orientation with respect to the vehicles. Additionally, the calibration process needs to account for the gravitational acceleration. This orientation problem has been extensively studied, please refer to [5] for details. The proposed skidding detection algorithm uses the acceleration vector (based on both vehicle's acceleration/deceleration and rotation) towards the vehicle's moving direction, denoted by $a(t)$ at time *t*. The raw data are passed through a low-pass filter (e.g., 2$^{nd}$ order Butterworth filter) to remove their noise component.

Step 2) Ground Speed Estimation

The instantaneous ground speed $v_{ground}(t_i)$ at time $t_i$ is calculated by integrating $a(t)$ within a sliding time window (will be explained in Step 3)) plus the wheel speed $v_{wheel}(t_0)$ at the beginning of that time window $t_0$. It is defined as:

$$v_{ground}(t_i) = v_{wheel}(t_0) + \int_{t_0}^{t_i} a(t) dt \quad (1)$$

More specifically, $v_{wheel}(t_0)$ serves as the "base" speed for the estimation, and the trapezoidal integration is used to calculate the speed change due to vehicle acceleration/deceleration.

Step 3) Skidding Detection

The time series of $v_{ground}(t_i)$ and $v_{wheel}(t_i)$ data are evaluated by a sliding time window. Let $T(t_0)$ denotes the sliding window with a length of *len* seconds, and $t_0$ indicates the beginning of the current time window being examined. The algorithm calculates the standard deviation of the set of absolute difference between the ground speeds and the wheels speeds in $T(t_0)$, which is defined as follows:

$$\sigma = std(\{abs(v_{ground}(t_i) - v_{wheel}(t_i)) \mid t_i \in T(t_0)\}) \quad (2)$$

where $\sigma$ is the standard deviation, *std* and *abs* are functions to calculate standard deviation and absolute value respectively.

If $\sigma > d_{skid}$, where $d_{skid}$ is a pre-defined threshold for vehicle skidding, the algorithm asserts a skidding event is detected.

Step 4) Advancing the Sliding Time Window

The sliding time window is advanced by *len* seconds if the vehicle is accelerating or decelerating. Acceleration and deceleration are monitored based on the average ground speed of the current time window. If the vehicle is moving in a constant speed or stopped, a smaller advancing step (1 second in our experiment) is applied in order to capture the changes of vehicle movements.

## V. EVALUATION

The goal of this evaluation is to demonstrate that VehSense is able to detect skidding events and monitor slippery road conditions effectively. A 2009 Honda Civic was equipped with a LG Nexus 5 smartphone and a Bluetooth OBD-II adapter. The proof-of-concept test consists of four trials of driving: one trial was conducted during normal driving condition, and three trials after about 6 inches of snow on unplowed road (refer to Fig. 3). Each trial was conducted on a different day that spans January, February 2016, at Buffalo, NY. Due to safety concerns, we tested our system in a parking lot and limited the vehicle speed to 20 mile per hour (about 9 meter per second). During the trials, the driver accelerates to 20 mph and then brings the vehicle to a full stop. The driver was asked to drive the car as in the normal conditions. During a stop-and-go process, if the overall handling of the vehicle has been affected, the driver will manually record the number of skidding events and their severity. We focused on vehicle skidding during acceleration and deceleration without directional or yaw changes.

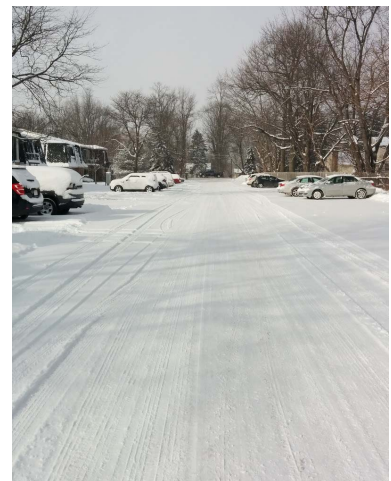

Figure 3. Testing Environment

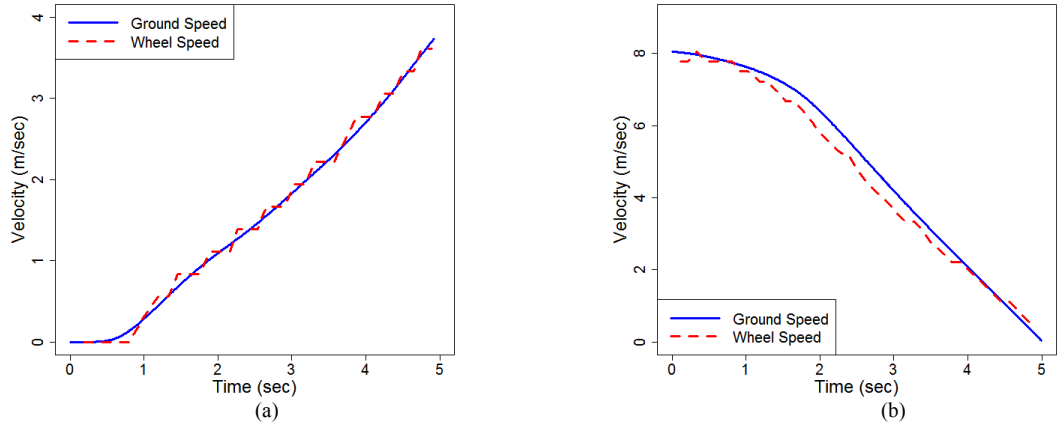

Figure 4. Speed comparison in typical normal driving conditions. (a) Normal acceleration. (b) Normal deceleration.

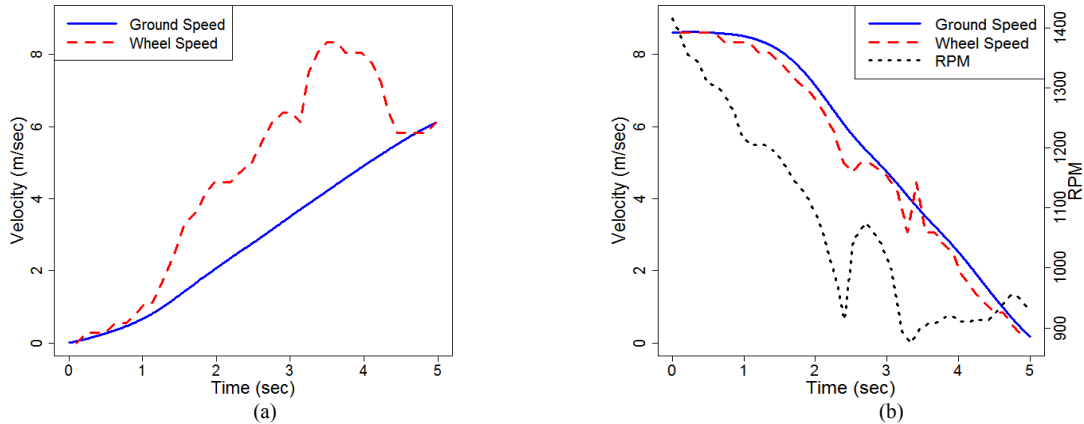

Figure 5. Speed comparison in typical vehicle skidding events. (a) Skidding during acceleration. (b) Skidding during deceleration.

Fig. 4 shows the ground speed and wheel speed during normal driving conditions. As we expected, there is a good match between ground speed and wheel speed during both vehicle acceleration and deceleration. The "staircase-shape" of the wheel speed is caused by the data type specified in the OBD-II standard: wheel speed is reported as an integer value.

Fig. 5 compares the ground speed and the wheel speed during typical skidding events. When vehicle is skidding, at around 1.5 sec. to 4 sec. in Fig. 5 a), there is a significant difference between ground speed and wheel speed, after the skidding event, wheel speed is in line with the ground speed. It shows that the proposed skidding detection algorithm is an effective approach. Skidding detection during deceleration is more challenging, as shown in Fig. 5 b), the variation between wheel speed and ground speed is notable but much less significant. This is caused by the electronic stability control and anti-lock brake systems in modern vehicles. One possible idea to improve the detector is to utilize vehicle RPM data, since such systems will result in engine output changes and gear shifts. We will explore this idea in future works.

The false negative and false positive rates are 10% and 2% respectively in our experiment, with $d_{skid} = 0.8$ and $len$ = 5 sec. There are a total of 30 stop-and-go processes during the snow trials, and 19 skidding events are recorded. The sliding window length $len$ and skidding threshold $d_{skid}$ are determined as follows: In order to cover the entire skidding event, $len$ is set to twice of the skidding duration (typical skidding lasted for about 2.5 seconds in our test). It is worth noting that as long as the sliding window covers most of a skidding event, VehSense is able to detect skidding events. This is because we are using the variations as the indicator, the exact speed difference will not affect too much of the result. The $d_{skid}$ is determined based on the variation retrieved from normal driving conditions, we used twice of the normal value as the skidding threshold.

This empirical method on choosing default parameters makes VehSense easy to deploy: after installation, VehSense will automatically determine the $d_{skid}$ by monitoring normal driving conditions, no action is required from the users. Different values of parameters have also been tested, as shown in Table I, the parameters determined by the empirical method provides good detection accuracy (although not always resulting in the minimum false negative and false positive rates).

TABLE I FALSE NEGATIVE/ FALSE POSITIVE RATES WITH DIFFERENT PARAMETERS

|  | $d_{skid} = 0.6$ | $d_{skid} = 0.7$ | $d_{skid} = 0.8$ | $d_{skid} = 0.9$ | $d_{skid} = 1$ |
|---|---|---|---|---|---|
| $len = 3$ | 6%/7% | 6%/4% | 6%/2% | 10%/1% | 31%/0% |
| $len = 4$ | 6%/5% | 11%/4% | 10%/2% | 10%/2% | 26%/0% |
| $len = 5$ | 0%/6% | 6%/4% | 10%/2% | 10%/2% | 26%/0% |
| $len = 6$ | 0%/11% | 11%/7% | 10%/5% | 10%/5% | 26%/0% |
| $len = 7$ | 0%/15% | 6%/11% | 6%/8% | 6%/5% | 26%/0% |

*Note: values are shown as False Negative Rate/False Positive Rate*

## VI. CONCLUSION AND FUTURE WORK

In this paper, we present the design, implementation, and evaluation of VehSense, a vehicular sensing platform that uses smartphones and OBD-II adapters for slippery road detection. The system utilizes standardized OBD-II commands and Android sensor readings, and thus is compatible with most of the passenger vehicles. It does not require user calibration and can be readily integrated into navigation applications. Our pilot study at Buffalo, NY, shows that the system is an effective skidding detector in snow days.

In the future, we want to add more capabilities such as integrating information from heterogeneous sources and supporting environmental sensors. More rigorous and larger scale experiments are also necessary. The objectives of the VehSense system are: 1) develop a low-cost crowdsourcing data collection system; 2) explore how to integrate this data with other contextual information and use it to detect slippery road, pavement roughness and other potential uses; and 3) evaluate how to inform driver about safety information without creating distractions.